\documentclass[useAMS,usenatbib]{mnras}
\usepackage{graphicx}
\usepackage{rotating}
\usepackage{aas_macros}
\usepackage{amsmath}
\usepackage{color}

\def\etal{{\it et al. }} 

\title[NGC~4546-UCD1] {An Extended Star Formation History in an Ultra Compact Dwarf}

\author[Norris \etal] {Mark A. Norris$^{1}$\thanks{norris@mpia.de}, Carlos G. Escudero$^{2,3,4}$, Favio R. Faifer$^{2,3,4}$, Sheila J. Kannappan$^{5}$, \newauthor
Juan Carlos Forte$^{4,6}$ \& Remco C. E. van den Bosch$^{1}$
 \\
 \\
 \\
  $^1$ Max Planck Institut f\"{u}r Astronomie, K\"{o}nigstuhl 17, D-69117, Heidelberg, Germany \\
  $^2$ Facultad de Cs. Astron\'omicas y Geof\'isicas, UNLP, Paseo del Bosque S/N, 1900 La Plata, Argentina\\ 
  $^3$ Instituto de Astrof\'isica de La Plata (CCT La Plata - CONICET - UNLP) \\
  $^4$ Consejo Nacional de Investigaciones Cient\'ificas y T\'ecnicas, Rivadavia 1917, C1033AAJ Ciudad Aut\'onoma de Buenos Aires, Argentina \\
  $^5$ Dept. of Physics and Astronomy UNC-Chapel Hill, CB 3255, Phillips Hall, Chapel Hill, NC 27599-3255, USA \\
  $^6$ Planetario ÒGalileo GalileiÓ, Secretar\'ia de Cultura, Ciudad Aut\'onoma de Buenos Aires, Argentina}
\begin{document}

\date{Accepted 2015 May 29. Received 2015 May 28; in original form 2015 April 2}

\pagerange{\pageref{firstpage}--\pageref{lastpage}} \pubyear{2015}

\maketitle

\label{firstpage}

\begin{abstract}
There has been significant controversy over the mechanisms responsible for forming compact stellar systems like 
ultra compact dwarfs (UCDs), with suggestions that UCDs are simply the high mass extension of the globular cluster 
(GC) population, or alternatively, the liberated nuclei of galaxies tidally stripped by larger companions. Definitive 
examples of UCDs formed by either route have been difficult to find, with only a handful of persuasive examples of 
stripped-nucleus type UCDs being known. In this paper we present very deep Gemini/GMOS spectroscopic observations 
of the suspected stripped nucleus UCD NGC~4546-UCD1 taken in good seeing conditions ($<$ 0.7\arcsec). With these 
data we examine the spatially resolved kinematics and star formation history of this unusual object. We find no evidence 
of a rise in the central velocity dispersion of the UCD, suggesting that this UCD lacks a massive central black hole like 
those found in some other compact stellar systems, a conclusion confirmed by detailed dynamical modelling. Finally 
we are able to use our extremely high signal to noise spectrum to detect a temporally extended star formation history 
for this UCD. We find that the UCD was forming stars since the earliest epochs until at least 1-2 Gyr ago. Taken together 
these observations confirm that NGC~4546-UCD1 is the remnant nucleus of a nucleated dwarf galaxy that was tidally 
destroyed by NGC~4546 within the last 1-2 Gyr.

\end{abstract}


\section{Introduction}

Since their discovery a decade and a half ago \citep{Hilker99,Drinkwater00} ultra-compact dwarfs \citep[UCDs:][]{Phillips01}
have proven to be highly enigmatic and controversial. Their structural parameters (size, mass, and central velocity 
dispersion) place them uneasily between star clusters, such as classical globular clusters (GCs), and bona-fide
galaxies. Their transitional location in scaling relations led to their exact origin being strongly debated, between
those who favoured their formation as the most massive star clusters \citep[see e.g.][]{Fellhauer02,Fellhauer05,
Bruens12,Mieske12}, and those that considered them the remnant nuclei of galaxies that have been tidally stripped 
\citep[see e.g.][]{Bekki03,Drinkwater03,Pfeffer13}.

However, in recent years a consensus has begun to emerge that as a group UCDs are a ``mixed bag", made up 
of both massive GCs and stripped galaxy nuclei \citep{Hilker06,Norris&Kannappan11,Chiboucas11,
daRocha11,Brodie11,AIMSSI,AIMSSII,Pfeffer14}. This view has developed because of the dual realisation that while most 
UCDs have properties consistent with those expected of GCs (including age, metallicity, $\alpha$-element enhancement)
and appear in numbers consistent with their being the bright end of the GC luminosity function \citep[see e.g.][]{Hilker06,
Norris&Kannappan11,Mieske12}, some UCDs are undeniable outliers, and have properties more consistent with a galactic 
origin (see below). Additional evidence for a stripped nucleus UCD population comes from theoretical simulations which 
indicate that stripping should be an effective method of creating UCD-like objects \citep{Bekki03,Pfeffer13}, and should 
be relatively common in a $\Lambda$CDM cosmology \citep[e.g.][]{Pfeffer14}, where minor merger events are common, 
particularly at early epochs.

Unfortunately, despite this emerging consensus very few UCDs are easily and definitively classifiable into either of 
the two groups, making it difficult to estimate the efficiency of stripped UCD formation with any accuracy. In part 
this is because of significant overlaps in the predicted properties of either type of object. But additionally, it is also due 
to the fact that those definitive signatures of UCD origin are observationally difficult to detect. For example, one of 
the clearest discriminants between massive GCs and stripped nuclei is to be found in the presence or absence of a 
central massive black hole. Many galaxies are observed to harbour a supermassive black hole (SMBH), 
with the occupation fraction increasing with galaxy mass until essentially all galaxies more massive than 
M$_\star$$\sim$10$^{10.5}$M$_\odot$ \citep[e.g][]{Decarli07,Miller15} are thought to host SMBHs. After tidal 
stripping of their host galaxies these SMBHs should remain within the central bound star clusters, now transformed 
into a UCD. However, while the presence of an SMBH is definitive proof that a particular UCD formed in a tidal 
stripping event, the converse is not true, as many lower mass galaxies are not found to have SMBHs 
\citep[e.g. M33 and NGC205;][]{Gebhardt01,Valluri05}. Regardless, to detect the signature of a black hole, even a 
relatively massive one, in a faint object with a size on the sky of $<$ 1\arcsec\,requires very large telescopes combined 
with either excellent natural seeing or adaptive optics (AO) observations. 

To date only two high spatial resolution spectroscopic observations of UCDs are in the literature. Making use of 
good natural seeing conditions \cite{Frank11} used IFU observations to study Fornax UCD3, finding 
weak rotation ($\sim$3 km/s) but no clear sign of either the presence of a large black hole or of dark matter (another 
clear signature of a galactic origin). In contrast, \cite{Seth14} examined the unusually dense UCD M60-UCD1 
\citep{Strader13} finding that it contained a black hole comprising 15$\%$ of its total mass. In this case high resolution 
AO assisted spectroscopy was key, the dynamical to stellar mass ratio of this UCD as measured in natural seeing 
data is not unusual \citep{Strader13,AIMSSII}. In fact, even when including the influence of the black hole,
\citet{Seth14} find that the mass-to-light ratio of M60-UCD1 is still consistent with that of massive GCs (although the stellar
M/L decreases by 40$\%$ compared to the no black hole model). 
This observation strongly hints that many more massive black holes in UCDs remain to be discovered. Many other 
massive UCDs do indeed display unusual dynamical mass to stellar mass ratios indicative of the likely presence of a 
large black hole \citep[e.g. S999 with M$_{\rm dyn}$/M$_\odot$ = 8.2,][]{Janz15}, but without higher spatial resolution 
data no definitive conclusions can be drawn.

One other conclusive proof of a galactic origin that until now has not been examined is whether or not a UCD 
had an extended star formation history (SFH). This approach has the benefit that while good signal to noise 
spectroscopy is required, very high spatial resolution (and consequently AO assisted) observations are not. Despite 
recent discoveries of multiple stellar populations within Milky Way GCs \citep[see e.g.][]{Gratton12}, GCs are still a 
very good approximation of idealised single stellar populations. UCDs which formed as giant globular clusters should 
be similar, or if they formed from the merger of several normal GCs in a dense star forming region, the age spread 
should at least be small \citep[$\la$ 200 Myr:][]{Fellhauer05}. The same is not true of nuclear star clusters (NSC), some of 
which are observed to have recent or ongoing star formation \citep{Rossa06,Walcher06,Seth06,Georgiev14}, 
indicating very extended formation times. Therefore, detecting an extended star formation history would be a true 
``smoking gun" of a stripped nucleus origin for a UCD.

In this paper we examine the UCD of NGC~4546 for signs that it is of galactic origin. NGC~4546 is a 
potentially fruitful place to look for a stripped nucleus UCD, because it is located in a lower-density environment (i.e. 
not a cluster core), where it is possible for gas-rich (and hence young) galaxies to undergo low-speed interactions of 
the type necessary to lead to tidal stripping. In galaxy clusters only old populations exist, making it difficult to separate 
stripped nucleus UCDs from the equally old GC population, while in the field it is possible to form stripped nucleus UCDs 
that are younger than the GC population of the associated galaxy. In \cite{Norris&Kannappan11} we have already 
shown that there are several good reasons to believe that NGC~4546-UCD1 is a former nucleus. Firstly, this UCD is 
$\sim$ 3 magnitudes brighter than the brightest GCs of NGC~4546, far too bright to be simply explained as the bright 
end of the GC luminosity function. Secondly, it is observed to counter-rotate its host galaxy, and in the same sense as 
a counter-rotating gas feature seen in the host galaxy. Thirdly, its alpha-element enhancement ratio [$\alpha$/Fe] is 
close to solar, indicating that the gas that formed it was enriched over an extended period, while most GCs are 
[$\alpha$/Fe] enhanced indicating very rapid star formation. Finally, it is young, around 3 Gyr, while its host galaxy is 
uniformly old ($>$ 10 Gyr). It is these facts, combined with its relative brightness (V = 17.6) and its relative closeness, 
only $\sim$13 Mpc (M-m = 30.58 $\pm$ 0.2) away, that make this object a prime candidate to search for signatures 
of a galactic origin for a UCD.

\section{Observations}

\subsection{Gemini/GMOS Spectroscopy}

The Gemini/GMOS \citep{Hook04} spectroscopy utilised in this paper was observed in the period July 2nd 2013 to January 
7th 2014 as part of Gemini program GS-2013A-Q-26. The main purpose of this program is to probe the connection, 
if any, between NGC~4546-UCD1 and the globular cluster system of NGC~4546. The results of this larger study will be 
presented in two forthcoming papers focussing on the photometric (Faifer et al. in prep) and spectroscopic properties 
(Escudero et al. in prep) of the UCD and GC system. In this paper we focus on the information provided by the spectrum 
of NGC~4546-UCD1.

The spectroscopy described here comprises 12 usable exposures, each of exposure time 1850s, observed with 
the B1200 grating with the UCD being observed through a 0.5\arcsec\, MOS slitlet. Four additional 
exposures were obtained but not used in this analysis because the instrument was significantly out of focus and 
the spectral resolution was a factor of $\sim$2 worse than desired. The resolution of the 12 good individual 
exposures varied from 1.26 to 1.57\AA\, FWHM as measured using isolated lines on the calibration arc spectra 
and the 5197\AA\, sky line which was visible on 11 of 12 of our science exposures. The spectral resolution measured 
from the same sky line on the summed 2D science frame was 1.41 $\pm$ 0.06\AA\, with the line being well fit by a 
Gaussian line spread function. The spectral setup utilised was sampled with 0.5\AA\, pixels over a wavelength range of 4080 to 
5575\AA. The seeing of the individual exposures varied from 0.5\arcsec\, to 0.9\arcsec\, with the median value being 
$\sim$0.7\arcsec as measured from the spatial distribution of the summed 2D spectrum of a confirmed Milky Way 
star (i.e. a point source). The alignment accuracy of the slitlets as measured from the alignment
images was around 0.1\arcsec, for all but one exposure which was offset by around 0.2\arcsec.

The reduction of the GMOS spectroscopy was initially similar to that outlined in \cite{Norris08}. The GEMINI IRAF 
package was used to carry out overscan subtraction, bias subtraction, flatfielding, extraction of the individual 
2D slitlets, wavelength calibration, and rectification. Further reduction was then achieved using a set of custom
IDL scripts. These scripts spatially rectified the 2D spectrum by tracing the peak of the light distribution, corrected
for heliocentric motion (extremely important as the heliocentric velocity varies from -30 to +29 km/s for this data set),
sky subtracted the individual exposures, and finally spatially aligned and coadded the individual exposures (with 
a clipping to reject the unexposed chip gaps and bad pixels). Finally, IRAF APALL was used to trace and extract a 
single integrated spectrum of the UCD, while an IDL script was used to extract a series of spectra spatially binned 
across the slit, each of which had a minimum signal-to-noise of 35 per \AA. 

Because of the relative brightness of the UCD compared to the GC targets on the same mask (the UCD is $\sim$3 
mag brighter), the S/N of the resulting integrated spectrum is excellent. The integrated spectrum has a S/N of 
between 50 per \AA\, at 4100\AA, and 200 per \AA\, at 5500\AA.

Flux calibration was achieved using an exposure of EG131 obtained as part of this science program using the
same observational setup as the main science data, except that the standard star was observed through a 0.5\arcsec\,
wide longslit, rather than the MOS mask. The reduction of the standard star data followed the same procedure 
as the science data.

\subsection{Gemini/GMOS Imaging}

As well as spectroscopic observations we have also obtained deep imaging of the environs of NGC~4546, primarily 
in order to select GCs for spectroscopic followup. However, this deep photometry is also extremely useful in estimating 
the stellar mass of NGC~4546-UCD1. The imaging was obtained with GMOS and comprises 4 $\times$ 100s exposures 
in the g',r',i' filters and 4 $\times$ 290s exposures in the z' filter \citep{Fukugita96}. GMOS imaging provides a field of 
view of 5.5\arcmin $\times$ 5.5\arcmin\, field-of-view, we used 2 $\times$ 2 binning, 
giving a pixel scale of 0.146 arcsec/pixel. The subexposures were dithered to remove the gaps between the 3 
GMOS CCDs and also to facilitate cosmic-ray removal.

The raw data were processed with the Gemini-GMOS IRAF package (e.g. gprepare, gbias, giflat, gireduce and gmosaic),
using appropriate bias and flat-field frames from the Gemini Science Archive (GSA).  As final step in the reduction 
process, the resulting images for each filter were aligned and co-added using the task imcoadd.

To perform the photometry of the UCD, we first modelled and subtracted the sky background and light from the halo of 
NGC~4546. To do this, we used an iterative combination of the SExtractor software \citep{Bertin_sextractor} and median 
filtering within IRAF \citep[see e.g.][]{Faifer11}. This procedure, provides a galaxy light subtracted image. Aperture 
photometry was then obtained for NGC\,4546-UCD1, using the PHOT task with a fixed 4.3 arcsec ($>$ 10R$_{\rm e}$ 
for this UCD) aperture. The derived magnitudes are entirely consistent with ones derived using an
alternative curve-of-growth analysis. Standard star fields, downloaded from the GSA, were used to achieve the transformation to the 
standard system. Finally, we applied the galactic extinction coefficients given by \citep{Schlafly11}, (A$_g$=0.112, A$_r$=0.077, A$_i$=0.057, 
A$_z$=0.043). Table \ref{tab:properties} gives the photometry for NGC\,4546-UCD1.

%----------------------------------------------------------------------------------------------------------------------
\begin{table}
\begin{center}
\caption[Properties]{Table of properties for NGC~4546-UCD1.}
\label{tab:properties}
\begin{tabular}{lll} \hline
Property				&	Value		 	\\
					&					\\
\hline
R.A. (J2000)			& 	 12:35:28.7		\\ 
Dec. (J2000)			& 	 --03:47:21.1		\\ 
Distance				& 	13.06 $\pm$ 1.26 Mpc				\\
					&						\\
\multicolumn{2}{c}{Photometry}&				\\
B					& 	18.57 $\pm$ 0.05 mag$^\star$	\\
V					& 	17.64 $\pm$ 0.04 mag$^\star$	\\
R					& 	17.04 $\pm$ 0.03 mag$^\star$	\\	
g$^\prime$			&	18.13 $\pm$ 0.01 mag	\\
r$^\prime$			& 	17.39 $\pm$ 0.01 mag	\\
i$^\prime$				& 	17.01 $\pm$ 0.01 mag	\\
z$^\prime$			& 	16.73 $\pm$ 0.05 mag	\\
J					& 	15.70 $\pm$ 0.11 mag$^\star$	\\
H					& 	14.69 $\pm$ 0.23 mag$^\star$	\\
K$_{\rm s}$			& 	14.86 $\pm$ 0.17 mag$^\star$	\\	
3.6$\mu$m			& 	14.40 $\pm$ 0.16 mag	\\	
					&						\\
\multicolumn{2}{c}{Derived Quantities}&				\\
R$_{\rm e}$			& 	25.54 $\pm$ 1.30 pc		\\
V					& 	1225.6 $\pm$ 0.3 km/s	\\
$\sigma$				& 	23.5 $\pm$ 2.5	km/s		\\
$\sigma_\infty$			& 	21.6 $\pm$ 2.5	km/s		\\
M$_{\star, \rm SED}$	& 	3.27$^{+0.85}_{-0.90}$$\times$10$^7$ M$_\odot$					\\
M$_{\star, 3.6\mu m}$	& 	3.11$^{+0.66}_{-0.66}$$\times$10$^7$ M$_\odot$					\\
M$_{\rm vir}$			& 	2.17$^{+0.51}_{-0.51}$$\times$10$^7$ M$_\odot$					\\
M$_{\rm dyn}$			& 	2.59$^{+0.49}_{-0.49}$$\times$10$^7$ M$_\odot$					\\

					&						\\
\multicolumn{2}{c}{Luminosity Weighted Stellar Populations}&				\\
Age					&	3.0 $\pm$ 0.05	Gyr			\\
$[$Z/H$]$				&	0.29 $\pm$ 0.01 dex			\\
$[\alpha$/Fe$]$			&	--0.02 $\pm$ 0.02 dex			\\

\hline
\end{tabular}
\end{center}
$^\star$The B, V, R, J, H, $\&$ K$_{\rm s}$ photometry is from \cite{Norris&Kannappan11}.
\end{table}
%----------------------------------------------------------------------------------------------------------------------

\subsection{Spitzer IRAC Imaging}
\label{Sec:IRACim}

Several papers have recently demonstrated that the WISE W1 or Spitzer IRAC 1 bands at 3.4/3.6$\mu$m are 
an almost ideal stellar mass tracer for older (age $>$ 2 Gyr) stellar populations \citep[see e.g.][]{Meidt14,WISEI,
Querejeta15}, and even for actively star forming disks \citep{McGaugh15}. Furthermore, as shown in Norris et al. (in prep) these photometric bands are remarkably insensitive 
to the detailed star formation history (SFH) of a stellar population, especially when combined with a spectroscopically 
derived luminosity-weighted age and metallicity for the stellar population. In such cases, the systematic uncertainty on
a mass-to-light ratio introduced by an imperfect knowledge of the SFH is always less than 15$\%$, when the luminosity 
weighted age is $>$ 2 Gyr, although systematic uncertainties due to stellar population synthesis modelling 
and assumed IMF can be considerably higher. Given these advantages we have examined archival Spitzer Space 
Telescope IRAC 1 (3.6$\mu$m) band imaging of NGC~4546 in order to derive a stellar mass for NGC~4546-UCD1 
that can be compared to estimates made using more common approaches (i.e. SED fitting).

The observations that we utilise were observed in July 2008 as part of Program 50630 (PI: van der Wolk), investigating 
the interstellar medium, star formation, and nuclei of the SAURON sample of early type galaxies. The exposure time
was 30s, and in our analysis we make use of the IRAC 1 map produced by the Spitzer Heritage 
Archive\footnote{http://sha.ipac.caltech.edu/applications/Spitzer/SHA}. 

The first step in our analysis is to subtract the bright halo of NGC~4546 which would compromise photometry of the UCD. 
To do that we make use of a multistep approach. First we run SExtractor \citep{Bertin_sextractor} in a very sensitive mode, 
with detection threshold of 1.5$\sigma$, and a very small background mesh value (16 pixels), in order to detect all sources, 
even those buried deep in the galaxy halo. The output of this analysis is an output image segmented into detected sources, 
this image is then converted into a pixel mask image, where every source is masked, other than NGC~4546 itself. We then 
use the IRAF task ELLIPSE in combination with the pixel mask to fit NGC~4546, with the resulting best fit model being 
subtracted from the original input to produce a galaxy subtracted image.  

Because the compact UCD (with R$_{\rm e}$ = 0.4\arcsec) is unresolved in IRAC 1 imaging (resolution $\sim$ 1.7\arcsec) 
we use aperture photometry to measure its integrated magnitude. Following the procedure outlined in \cite{Reach05}, we
use a 10 pixel aperture (1 pixel = 0.6\arcsec) and a sky annulus of 12 to 20 pixels. Using the conversions provided in 
\cite{Reach05} from MJy to magnitudes, we derive a total integrated magnitude in the IRAC 1 band of 14.40 $\pm$ 0.16 mag
(where we have added a conservative error of 0.15 mag in quadrature to the photometric error, to account for the poorly 
constrained error due to the galaxy background and IRAC scattered light) for the UCD.

\subsection{Additional Data}
In addition to the new data described above, our analysis also makes use of SOAR/Goodman optical imaging in the B, 
V, and R bands, and 2MASS J, H, and K$_{\rm s}$ imaging. The reduction and analysis of both sets of data was described 
in \cite{Norris&Kannappan11}. We also utilise the only available HST imaging of NGC~4546, a single WFPC2 image in
the F606W filter, the analysis of which was also described in \cite{Norris&Kannappan11}.

\section{Results}

\subsection{Kinematics}

We use the penalised pixel fitting code pPXF \citep{ppxf} to measure the UCD kinematics. As in \cite{AIMSSI} we make 
use of the stellar population synthesis models of \cite{MarastonStromback11} as the template spectra. In particular,
we use the high resolution (0.55\AA\, FWHM) ELODIE-based SSP models, because they are currently the only 
comprehensive set of SSP models (as opposed to libraries of individual stellar spectra) available with resolution higher 
than our spectra (1.41\AA).

\subsubsection{Integrated Kinematics}

\begin{figure*} %  figure placement: here, top, bottom, or page
   \centering
   \begin{turn}{0}
   \includegraphics[scale=1.0]{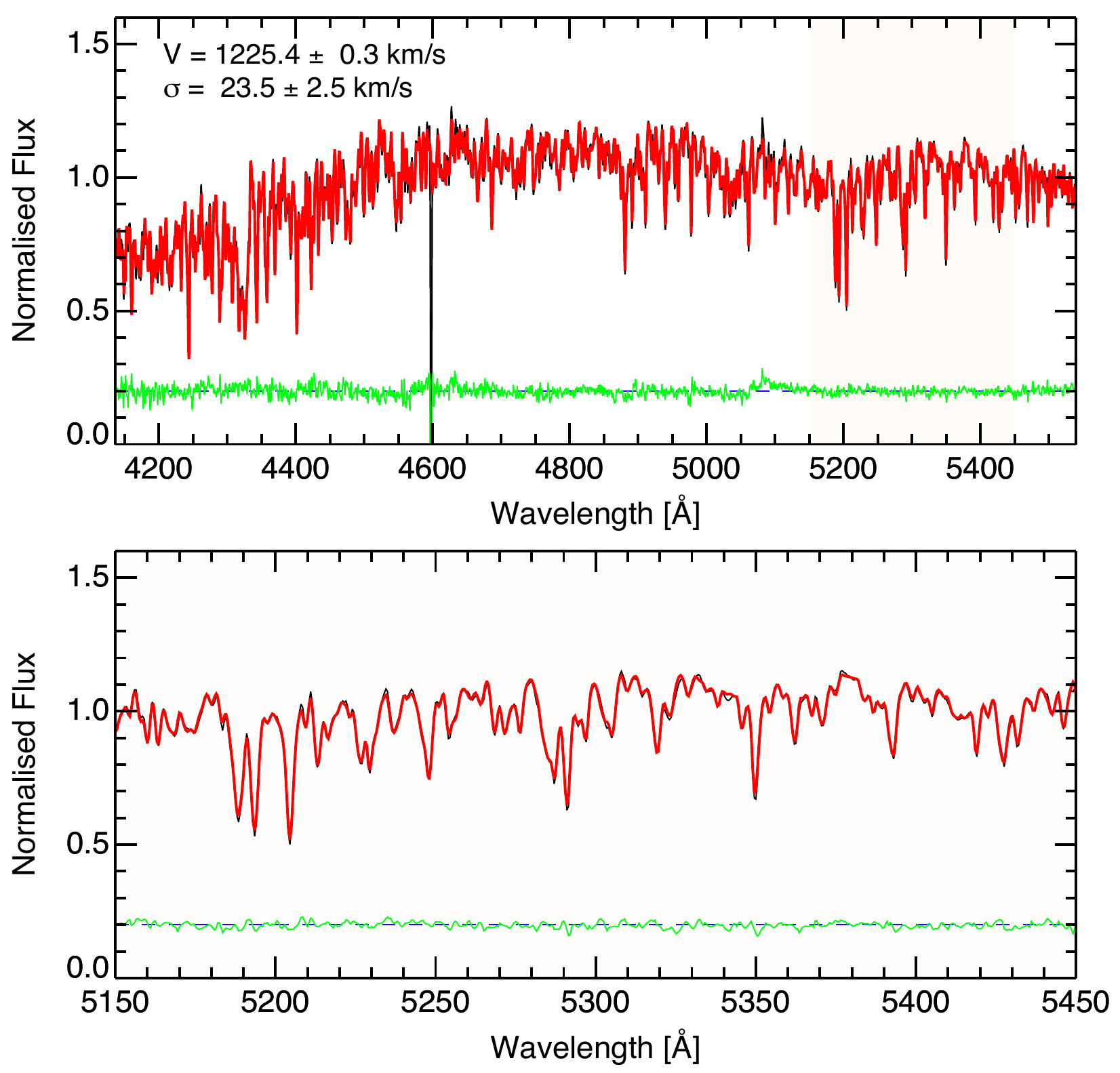}
   \end{turn} 
   \caption{$\bf{Upper~Panel:}$ Integrated spectrum of NGC~4546-UCD1. The black line is the spectrum, 
   the red line is the best fit found by pPXF using the models of \citet{MarastonStromback11}. The best fit 
   kinematics are provided in the upper left. The green line shows the residuals (offset by 0.2 to make them 
   visible), the blue dashed line behind the green line shows the no residuals line for a perfect fit. The residual
   features at $\sim$ 4600  and 5080\AA\, are due to chip gaps and are masked during the fitting. The shaded
   region indicates the region displayed in the lower panel.
   $\bf{Lower~Panel:}$ Zoom in of the region between 5150 and 5450\AA\,to demonstrate the high quality of
   the fit.
   }
   \label{fig:integrated_kins}
\end{figure*}

We first examine the kinematics of the integrated spectrum of NGC~4546-UCD1. We fit the integrated spectrum over
the wavelength range 4120 to 5530\AA, masking two regions affected by residuals from chip gaps (around 4600 and 5080\AA). 
We first attempted fitting the first four moments of the line-of-sight velocity distribution, however, the higher order moments 
h$_3$ and h$_4$ were consistent with zero and so we refit using only the first two moments, the velocity and velocity dispersion. 
We estimate the uncertainties on the measured parameters using 100 Monte Carlo resimulations of the data 
with the measured uncertainties on the spectra and the spectral resolution. Figure \ref{fig:integrated_kins} shows the result 
of the fitting procedure. It is clear that the fit is excellent, with only slight mismatch between the template and the model 
in the region where problems due to the chip gaps remain. The excellent fit also indicates that as expected there is no 
significant H$\beta$ or [OIII] emission present in the UCD.

The measured recessional velocity is in good agreement with that measured from the previously presented lower-resolution 
SOAR/Goodman spectroscopy \citep{Norris&Kannappan11}, and also with the remeasured velocity derived 
from the higher resolution SOAR/Goodman spectrum presented in \cite{AIMSSI} after this spectrum was reanalysed using 
the same method as here\footnote{The previous estimation used IRAFs FXCOR to derive the velocity, and included an incorrect
sign for the heliocentric correction for this object leading to an apparent offset in velocity.}. It is interesting that the measurement 
of velocity dispersion determined in \citet{AIMSSI} (21.8 $\pm$ 2.5) and the one determine here (23.5 $\pm$ 2.5)  agree so well, 
despite the significantly improved seeing available in the observations with GMOS ($\sim$0.7\arcsec vs $\sim$1.4\arcsec). 
This is suggestive that the velocity dispersion of NGC~4546-UCD1 is not strongly centrally peaked.

\subsubsection{Spatially Resolved Kinematics}
 \label{Sec:Res_Kins}
 
 \begin{figure} %  figure placement: here, top, bottom, or page
   \centering
   \begin{turn}{0}
   \includegraphics[scale=0.95]{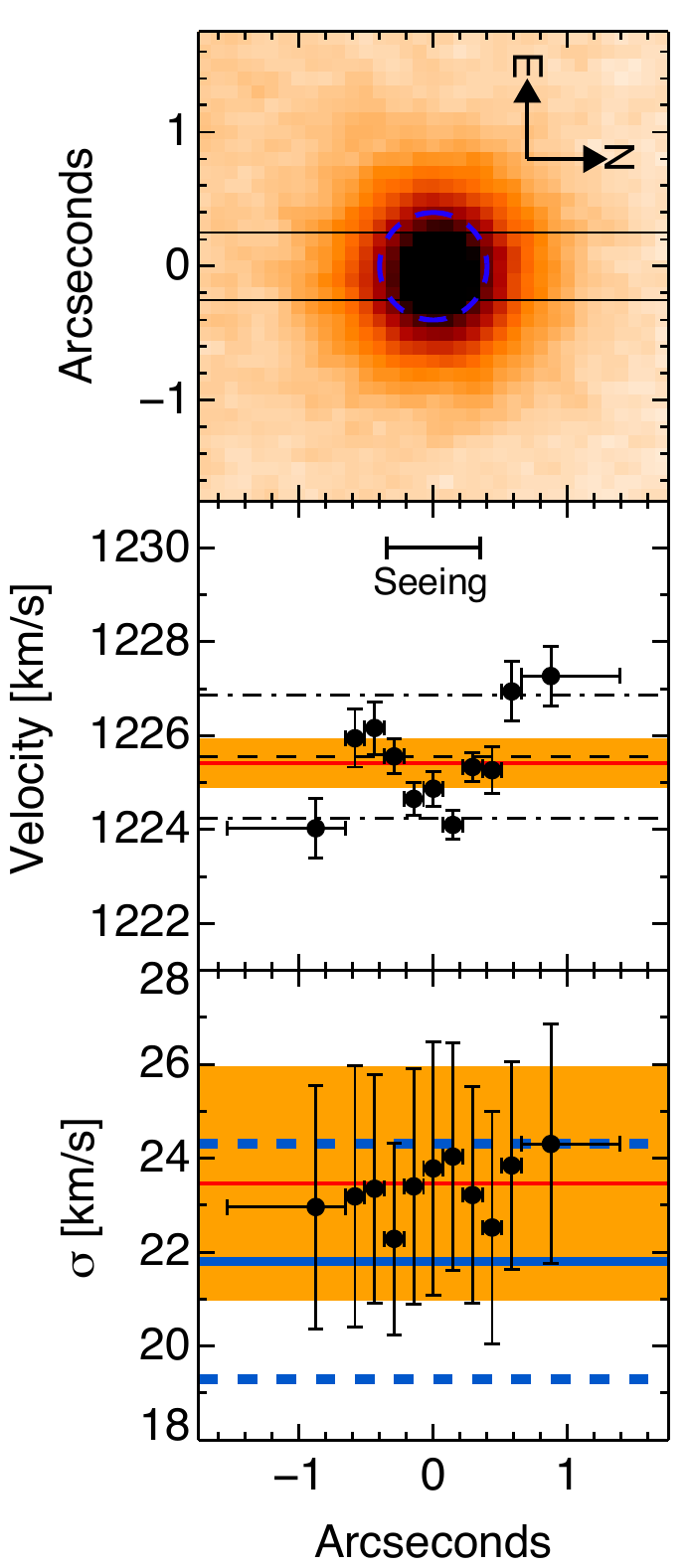}
   \end{turn} 
   \caption{$\bf{Upper~Panel:}$ Thumbnail image (3.5\arcsec$\times$3.5\arcsec) of NGC~4546-UCD1 from 
   HST WFPC2 F606W imaging. The black lines indicate the size and orientation of the 0.5\arcsec\,slit
   used in this study. The blue dashed circle indicates the radius encompassing half
   the flux of NGC~4546-UCD1.
   $\bf{Middle~Panel:}$ The spatially resolved velocity of NGC~4546-UCD1. The x-axis error bars
   show the lower and upper limits of the binned spectra, the dot shows the light weighted centre
   of each bin. The red line and orange bands show the 
   recessional velocity and its 1-$\sigma$ uncertainties as measured from the integrated spectrum.
   The dashed line shows the most probable central velocity, as fit to the spatially resolved velocities, 
   and the dot-dashed lines the most probable maximum and minimum velocities of the rotation curve. 
   The most probable amplitude of rotation is 1.3 km/s.
   The integrated seeing of 0.7\arcsec\, is indicated at the top of the panel.
   $\bf{Lower~Panel:}$ The velocity dispersion profile of the UCD. 
   The red line and orange region again shows the velocity dispersion and its 1-$\sigma$ uncertainty 
   as measured from the integrated spectrum. The blue solid line and the blue dashed lines show the 
   velocity dispersion and its 1-$\sigma$ error measured in \citet{AIMSSI} using a spectrum
   from the SOAR telescope. The velocity dispersion is consistent with being flat over the full extent 
   of the UCD.}
   \label{fig:resolved_kins}
\end{figure}

We next turn to our attempt to extract spatially resolved kinematics from our spectroscopy. As
described previously we spatially binned our spectrum along the slit to a minimum signal-to-noise of
35 per \AA\,(as measured in the region 5000 to 5050\AA), this yielded 11 separate spectra
across the UCD, with the central 9 pixels all reaching that S/N with a single spatial pixel.
We then fit each spectrum using exactly the same procedure outline above.

Figure \ref{fig:resolved_kins} shows the result of the fitting procedure. We are able to detect
rotation of amplitude 1.3 km/s (as measured using a probabilistic min/max procedure to 
determine the most likely lower and upper rotation velocity) across the UCD, with a central
recessional velocity that is very consistent with that derived from the integrated spectrum. 

The small amplitude of rotation is close to the limit that our spectroscopic setup would be
expected to reliably recover, we have therefore examined how robust this detection is to
changes in spatial binning, wavelength range fit, and number of LOSVD moments fit. In part 
we do this to attempt to probe the effect of systematic uncertainties such as template mismatch 
etc. We find that while the exact amplitude of the rotation varies between 1 and 2 km/s, the 
detection of rotation, and its sign (i.e. which side recedes or approaches) is a robust result of 
the fitting procedure.

The general shape of the rotation curve is also intriguing, as the apparent reversal in direction
around -0.5\arcsec\,is also a robust result of the kinematic fitting procedure. The shape itself
could be indicative of disturbed kinematics, such as a counter rotating stellar component. These
features, known as kinematically decoupled cores have been observed in a range of galaxy 
types from massive early types \citep[e.g.][]{SAURONIX} to low mass dwarf ellipticals where 
they typically display both counter-rotation and younger stellar populations \citep{Toloba14}. 
In addition, complex counter rotating structures have been uncovered in several nuclear star 
clusters \citep{Seth10,Lyubenova13}, where again one rotating component is typically observed 
to be younger than the other population. Only higher spatial and spectral resolution data ideally 
from IFU spectroscopy will be able to confirm the significance of this feature.

It is important to note that the rotation amplitude measured is only a lower limit as the slitlet 
was arbitrarily aligned with the UCD and likely misses the direction of peak rotation. In fact the 
position angle of this UCD, which is slightly flattened (b/a = 0.92), is around 82$^\circ$, which 
means that the slitlet was almost perfectly aligned with the minor axis of the UCD, and further
supports the possibility of significant rotation being present but unobservable with this dataset. It is not 
clear what rotation amplitude, if any, a UCD should be expected to have, as only two previous 
measurements have been made. Typical globular clusters do indeed rotate with a few km/s
\citep{Bianchini13,Lardo15,Kimmig15}, as does UCD3 in the Fornax cluster \citep{Frank11}, 
while the nuclear star cluster of NGC~4244 rotates with an amplitude of 
$\sim$ 30 km/s \citep{Seth08} and the unusually dense UCD, M60-UCD1, is observed to rotate 
with amplitude as high as 40 km/s \citep{Seth14}. A rotation amplitude of this magnitude would 
be clearly visible in our data, except that the slit may be aligned almost perfectly 
along the zero velocity contour.
 
In contrast with the rotation curve, the velocity dispersion profile of NGC~4546-UCD1 is
consistent with being flat, indicating that we either lack sufficient resolution, or that unlike 
M60-UCD1, this UCD does not host a massive black hole at its centre.

\subsection{Luminosity-Weighted Stellar Populations}
\label{Sec:LICK}

In order to compare the integrated stellar population parameters of NGC~4546-UCD1 to previous
lower S/N measurements for the same object, and to literature measurements of globular clusters,
we undertake a simple Lick/IDS line strength analysis \citep{Worthey97,Trager98}. Due to the high S/N
of our input spectrum the statistical uncertainties on this analysis are unreasonably small. Therefore,
to derive more realistic errors on the derived stellar population parameters we attempt to determine their
systematic uncertainties by using several different sets of SSP models.

Firstly we use the method outlined in \cite{Norris08}; the Lick/IDS line strengths are measured 
from the flux calibrated GMOS spectrum, then we apply small offsets in order to bring the GMOS 
indices into better agreement with the Lick/IDS system \citep[see][for more details]{NSK06,Norris08}.
We then use the procedure outlined in \cite{Norris&Kannappan11} to convert our measured Lick/IDS
indices into integrated stellar population parameters for the UCD. In short, we use the $\chi^2$-minimization 
approach of \cite{Proctor04}. To ensure a smooth model grid we first interpolate the SSP models 
of \citep{Thomas03,Thomas04} to a finer model grid, and then we perform a $\chi^2$-minimization 
on the resulting age, [Z/H] and [$\alpha$/Fe] space. Errors come from 50 Monte Carlo resimulations of the 
input data within the measured index errors.

Secondly, we utilise the same method outlined above except that we make use of 
the latest version of the \citet{Thomas11} models. In this version of the models no offset to the Lick/IDS 
system is required, as the models are computed for flux calibrated spectra such as ours. In addition, 
two flavours of model are provided for two sets of isochrones. We therefore fit both for both sets of
isochrones. We take the resulting fits from the Padova isochrone based models as our best fit
SSP-equivalent stellar populations.

We then estimate the uncertainties on the derived SSP-equivalent stellar population 
parameters by looking at the scatter in the derived stellar populations from the three available model sets 
(the single \citealt{Thomas03,Thomas04} model set, and the two \citealt{Thomas11} model sets).

Using the first procedure outlined above, with a more restrictive set of available indices measured from the SAURON
spectrum of the UCD, \cite{Norris&Kannappan11} found age = 3.4$^{+1.7}_{-1.2}$ Gyr, [Z/H] = 0.21 $\pm$ 0.14 
and [$\alpha$/Fe] = --0.01 $\pm$ 0.08. Using our high S/N, wider wavelength coverage spectrum we
find that the UCD has age = 3.99$^{+0.93}_{-0.75}$ Gyr, [Z/H] = 0.18 $\pm$ 0.06, and [$\alpha$/Fe] 
= 0.05 $\pm$ 0.05, values entirely consistent with those derived previously.

\subsection{The Star Formation History of NGC~4546-UCD1}

Because the integrated spectrum of NGC~4546-UCD1 is of such high quality (S/N of 50 to 200) we
are also able to examine the temporally resolved star formation history of the UCD. To do this we make use
of a similar approach to that outlined in \cite{McDermid15}. We again use the pPXF code used previously
to derive the UCD kinematics, in this instance we make use of the regularisation feature to fit a linear 
combination of SSP models to our UCD spectra. Unfortunately, due to the very high luminosity-weighted metallicity
of NGC~4546-UCD1 derived in the previous section, we are unable to use the \cite{MarastonStromback11} 
high resolution models used in the kinematic analysis, as these only extend to [Z/H] = 0.3 dex and
this leads to artefacts with the solutions running into the edge of the available model parameter space.
Likewise we are also unable to use the MIUSCAT SSP models of \cite{MIUSCATI}, which have been successfully 
utilised to determine the star formation histories of early type galaxies \citep[e.g.][]{McDermid15}, as they 
similarly do not extend to sufficiently high metallicity (maximum [Z/H]  = 0.22 dex). We therefore make use of 
the newest MILES based models from \cite{MILESII}. These models extend to [Z/H] = 0.4 dex for their BaSTI 
isochrone based version, and furthermore provide a choice of IMF and alpha-enhancement ratio. For this work we use 
the revised Kroupa IMF based models (see \citealt{MILESII} for full details), and assume solar-alpha enhancement 
in order to properly account for the measured [$\alpha$/Fe] of NGC~4546-UCD1. The chosen models span a 
range in age of 0.03 to 14 Gyr, and [Z/H] from --2.27 to +0.4 dex, providing a total of 636 models in all. As the 
resolution of the MILES models is lower than our spectra (FWHM = 2.51\AA\,vs 1.41\AA) we are forced to smooth 
our input spectra in order to match the lower resolution of the templates.

The regularisation prescription used by pPXF is designed to reduce the degeneracies present when
fitting star formation histories to the integrated spectra of complex stellar systems. It enforces the situation 
that adjacent models in the grid must have weights that vary smoothly, with the result that the fitted 
star formation history is the smoothest solution that is consistent with the noise. Despite this requirement,
bursty star formation histories are not prevented, only that the burst will appear somewhat broadened
to include adjacent models. As the SSP models used here are normalised to an initial birth mass of 1
solar mass the distribution of pPXF weights recovers the contribution of each SSP to the zero-age
mass distribution, which is equivalent to the star formation history of the population. 

\begin{figure*} %  figure placement: here, top, bottom, or page
   \centering
   \begin{turn}{0}
   \includegraphics[scale=1.05]{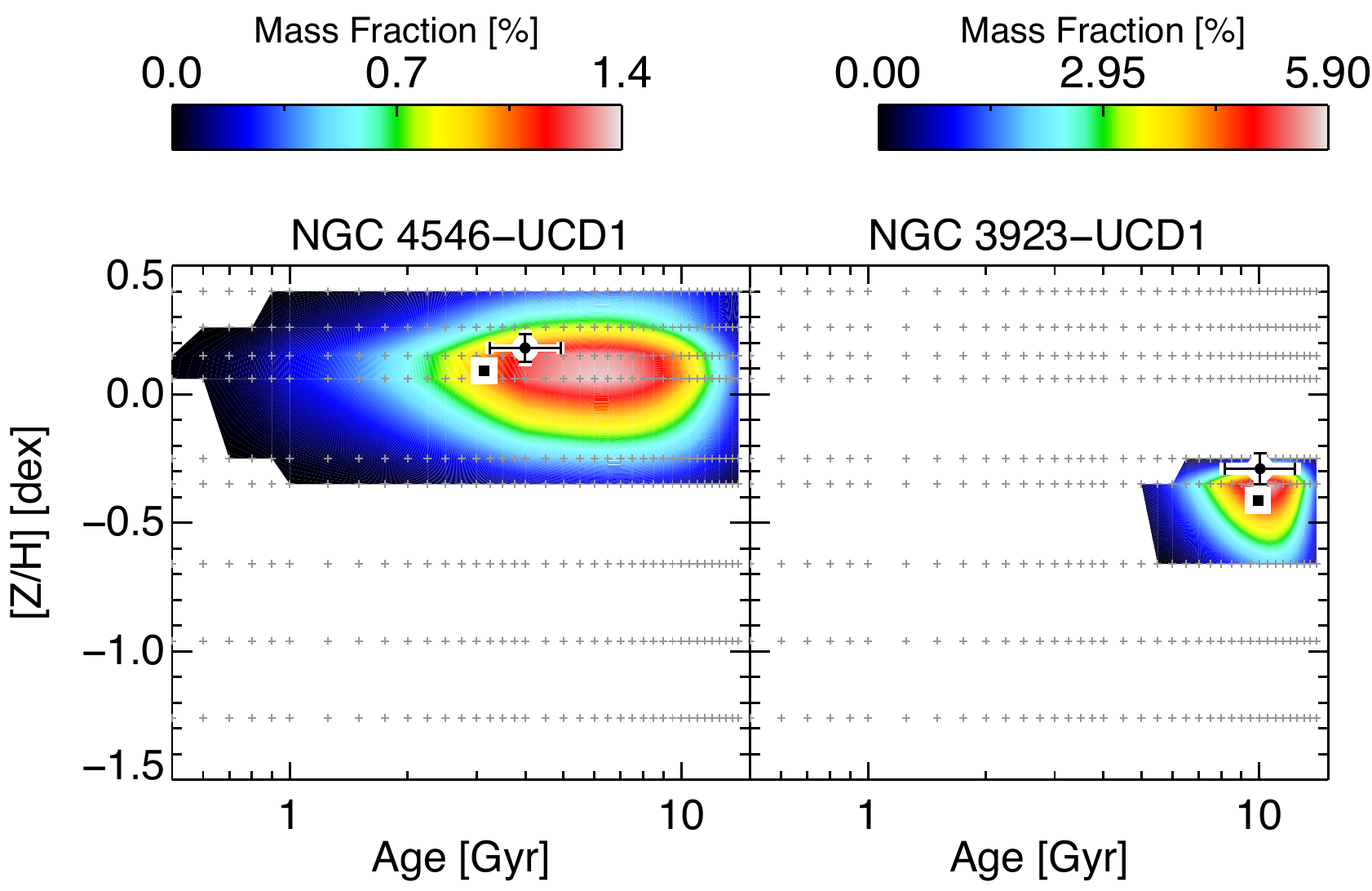}
   \end{turn} 
   \caption{ $\bf{Left~Panel}$ The star formation history found by fitting the integrated spectrum of NGC~4546-UCD1,
   with the SSP models of \citet{MILESII} using pPXF. The coloured contours indicate the weight of each SSP model 
   which is equivalent to the zero-age mass distribution of the SFH, white regions indicate models which contribute no flux to the best fit model. 
   The small grey plus symbols indicate the age and [Z/H] of each of the available models.  The white circle with the 
   black dot indicates the location of the luminosity weighted SSP age and metallicity 
   of this object as measured using absorption line strength indices (Section \ref{Sec:LICK}). 
   The black and white square shows the location of the luminosity-weighted age and metallicity
   derived from the full spectral fitting.
   $\bf{Right~Panel}$ The star formation history of the suspected massive GC-type UCD NGC~3923-UCD1.
   The black and white dots and squares are the same as in the left panel.
   }
   \label{fig:resolved_sfh}
\end{figure*}

Figure \ref{fig:resolved_sfh} shows the resulting star formation histories derived using this procedure.
As well as NGC~4546-UCD1, we also fit the star formation history of the suspected GC-type UCD
NGC~3923-UCD1. The spectrum of NGC~3923-UCD1 was obtained with GMOS using exactly the 
same setup as the spectrum of NGC~4546-UCD1, it was first presented in \cite{Norris12} and provides
a perfect comparison spectrum. The analysis of the luminosity weighted age and metallicity of 
NGC~3923-UCD1 (plus the additional UCDs of NGC~3923) will be presented in Janz et al. (in prep), 
where the analysis follows a similar method to that taken in Section \ref{Sec:LICK}. The derived age
of NGC~3923-UCD1 is 10.05$^{+2.28}_{-1.87}$ Gyr, its metallicity is [Z/H] = --0.29 $\pm$ 0.06, and
[$\alpha$/Fe] = 0.24 $\pm$ 0.04, meaning this object has properties entirely consistent with a typical
GC. 
 
The white and black circles and squares in Figure \ref{fig:resolved_sfh} display the luminosity-weighted
age and metallicity of the UCDs, with the circles indicating the luminosity-weighted populations measured
from the line-index based method described above (age = 3.99 Gyr, [Z/H] = 0.18 dex, and age = 10.05 Gyr, 
[Z/H] = --0.29 dex for NGC~4546-UCD1 and NGC~3923-UCD1 respectively). The squares indicate the luminosity-weighted 
populations as derived from the full spectral fitting method (age = 3.13 Gyr, [Z/H] = 0.09 dex, and age = 9.91 Gyr, 
[Z/H] = --0.41dex for NGC~4546-UCD1 and NGC~3923-UCD1 respectively). The fact that both methods provide
consistent results provides additional confirmation of the reliability of the full spectral fitting approach.

From Figure \ref{fig:resolved_sfh} it is clear that NGC~4546-UCD1 appears to have had an extremely 
extended star formation history, with stars forming from the earliest periods up until around
1 Gyr ago (as strongly hinted at by the young luminosity weighted age of the UCD). In order
to test how robust this finding is we have attempted to examine how many separate star formation
events would be required to produce an observed SFH like this one. To do this we created single, dual,
and triple burst SFHs using the same \cite{MILESII} models used to fit the data. The models had
metallicity of 0.06, 0.11 or 0.15 dex (as we are attempting to reproduce the SFH shown in Figure 
\ref{fig:resolved_sfh}, not the integrated properties of the UCD), and to keep the problem tractable 
each burst was assumed to have equal weight in the final light (i.e. later bursts formed fewer stars). 
The resulting spectra were then
degraded to have the same S/N as the spectrum of NGC~4546-UCD1 (i.e. $\sim$200\AA) and then
were fit with pPXF in the same manner as the real spectrum. The result of this procedure is that 
few models could reproduce the integrated properties of the UCD, and no single or double burst was 
able to reliably reproduce the star formation history seen in Figure \ref{fig:resolved_sfh}, in the sense
that they could not reproduce either the distribution of ages and metallicities, or the relative weights
of each model observed in Figure \ref{fig:resolved_sfh}. However, a triple 
burst model with bursts at 3, 6, and 10 Gyr and fixed metallicity of 0.11 dex did do a reasonable job of 
approximating the best fit SFH, and of matching the integrated properties of the UCD. We therefore 
conclude that the UCD has had an extended star formation history, but cannot determine if this history 
was continuous, bursty, or a combination of both modes.

The metallicity evolution indicated by Figure \ref{fig:resolved_sfh} is also particularly 
interesting, as at the earliest epoch the metallicity of the object was already high, at least near solar 
values. This high initial metallicity at early times implies that this object formed in an already highly enriched 
medium, which in turn implies that the object was most likely located in the very centre of 
a fairly massive halo in the early Universe. The high absolute value of the metallicity
is also unusual for most galaxy populations; only 38 out of 259 galaxies from the ATLAS$^{\rm 3D}$
have [Z/H] $>$ 0.15 within R$_{\rm e}$/8. This means that this UCD is as metal rich as the inner regions
of some of the most massive galaxies. Such high metallicities do however appear to be a fairly common
property of suspected stripped-nucleus UCDs (Janz et al. in prep); in particular the confirmed stripped-nucleus UCD
M60-UCD1 has an almost identical total metallicity of around 0.19, although with a lower [Fe/H] of around 
solar \citep{Strader13}.

The subsequent metallicity evolution of the NGC~4546-UCD1 is also intriguing. 
The fact that the metallicity remained fairly constant during the period of star formation indicates 
that either fresh gas or lower metallicity stars from outside the UCD was regularly accreted onto 
the proto-UCD, presumably from a gas-rich and star forming galactic disc. The accretion of fresh 
gas or stars from further out in the original galaxy may also explain the observed kinematic features 
described in Section \ref{Sec:Res_Kins}, as any accreted gas or stars would likely form dynamically 
distinct structures. 

It is noteworthy that the star formation history of NGC~4546-UCD1 is remarkably similar to those 
found for a sample of seven low-redshift massive compact galaxies by \citet{Ferre-Mateu12}. In
particular, the high metallicity and relatively large fraction of stars formed recently in both the
galaxies and NGC~4546-UCD1 is unusual in most low redshift early type populations, perhaps hinting 
at commonalities between the formation mechanism of NGC~4546-UCD1 and these higher mass 
objects. It is also intriguing that the implied extended star formation history (but not the 
metallicity enrichment history) of NGC~4546-UCD1, which was extended over $\sim$10Gyr, and 
potentially composed of a series of short bursts, looks remarkably similar to that observed for the 
presumed nucleus (the GC M54) and superposed core of the Sagittarius dwarf galaxy \citep{Siegel07}. 
In Sagittarius, the core + M54 region has a SFH that involved several distinct bursts over at least 10 Gyr, 
with the last star formation possibly occurring within the last Gyr.

In contrast, the behaviour of the suspected GC-type UCD NGC~3923-UCD1 is exactly 
as expected for an object formed in a single burst with a fixed metallicity. The model fits
are in excellent agreement with the age and metallicity derived from the Lick analysis,
and the spread in age and metallicity is entirely explainable due to model degeneracies
and fitting errors.

\subsection{Stellar Mass}
\label{Sec:stellarmass}

We make use of two independent approaches to estimate the stellar mass of NGC~4546-UCD1.

Firstly, as in \citep{AIMSSI}, we use a modified version of the stellar mass estimation code presented in \cite{Kannappan07} 
and subsequently updated in \citet[][first model grid described therein]{Kannappan13}. This code fits photometry from the 
Johnson-Cousins, Sloan, and 2MASS systems with an extensive grid of models from \citet{BruzualCharlot} assuming a 
Salpeter initial mass function (IMF), to be rescaled as will be described shortly. 
The input photometry (provided in Table \ref{tab:properties}) is fitted by a grid of two-SSP, composite old + young models 
with ages from 5~Myr to 13.5~Gyr and metallicities from Z = 0.008 to 0.05. The two component nature of this fitting 
procedure is ideally suited to a determination of a complex stellar population such as that spectroscopically observed in 
NGC~4546-UCD1. This is because the two component nature of the fitting procedure provides a reasonable 
approximation of an extended SFH, allowing the accurate recovery of the true integrated M/L. The stellar mass is determined 
by the median and 68$\%$ confidence interval of the mass likelihood distribution binned over the grid of models. 

We rescale the derived stellar masses by a factor of 0.7 in order to match the ``diet" Salpeter IMF of \citet{Bell01}. This 
is done, both to bring the stellar mass estimates from this technique into better agreement with estimates (such as the 
second approach provided below) made assuming a Chabrier or Kroupa type IMF, and because such IMFs appear to be 
a better fit to observational data than Salpeter for both GCs \citep{Strader11a} and relatively low mass early type galaxies 
\citep[those with $\sigma_{\rm e}$ $\sim$ 90 kms$^{-1}$;][]{ATLAS3DXX}. Using this approach we measure a stellar mass 
of 3.27$^{+0.85}_{-0.90}$$\times$10$^7$ M$_\odot$.

Secondly, we make use of the newly derived near infrared M/L ratios presented in \cite{WISEI} and Norris et al. (in prep) in 
combination with the luminosity-weighted stellar population parameters measured in Section \ref{Sec:LICK}. This 
approach has the advantage that the reduced sensitivity of the M/L to stellar population age and metallicity in the near IR 
\citep{Meidt14,WISEI,Roeck15} translates directly into a reduced sensitivity to SFH (Norris et al. in prep). In fact, as demonstrated 
in Norris et al. in prep, where a luminosity-weighted (i.e. SSP equivalent) age and metallicity are available for a composite stellar 
population with SSP age $>$ 2 Gyr, the SSP equivalent M/L is always within 10$\%$ of the true integrated M/L, irrespective of the 
SFH. For the UCD's measured age of 3.99 Gyr and [Z/H] of 0.18 dex, the M/L models of Norris et al. (in prep) predict a mass-to-light
ratio of 0.53$^{+0.05}_{-0.05}$ in the IRAC 3.6 $\mu$m band. We convert the measured IRAC 3.6 $\mu$m magnitude derived in 
Section \ref{Sec:IRACim} (14.4 mag) to total luminosity using the absolute magnitude of the sun M$_{\odot}^{3.6\mu m}$ = 3.24 \citep{Oh08} 
and an assumed distance modulus to NGC~4546 of 30.58 $\pm$ 0.2. This conversion leads to a total luminosity in the 3.6 $\mu$m 
band of 5.87$^{+1.51}_{-1.51}$$\times$10$^7$ L$_\odot$, where most of the uncertainty arises from the relatively poorly constrained 
distance to NGC~4546. This luminosity when combined with the appropriate M/L yields a stellar mass of 
3.11$^{+0.66}_{-0.66}$$\times$10$^7$ M$_\odot$.

\subsection{Dynamical Modelling}

In this section we use two approaches to estimate the total dynamical mass of NGC~4546-UCD1 and also
search for evidence of the existence of a supermassive black hole within this UCD.

\subsubsection{Virial Mass}

We first estimate the dynamical mass of NGC~4546-UCD1 using the virial equation:

\begin{eqnarray}
M_{\rm vir} = \frac{C \sigma^{\rm 2} R}{G}
\end{eqnarray}

where C is the virial coefficient, R is a measure of the size of the system, and $\sigma$ an estimate 
of the system's velocity dispersion. The exact value of C varies dependent on the structure of the
object, but using equation 11 from \cite{Bertin02} we are able to convert the measured Sersic n value
for this UCD (1.4) into a virial coefficient of 7.8. Note that this value of C is larger than the fixed value 
of 6.5 adopted by \citet{AIMSSII} in their analysis of a sample of GCs, UCDs and cEs including
NGC~4546-UCD1. This different choice of value of C is responsible for most of the difference between
our derived virial mass and theirs. We use the measured effective radius of 25.5 pc as the measure of the size of the system. 

The measured velocity dispersion must then be adjusted to take into account the fact that our slit has 
a finite width, and that this width is of similar size to both the seeing and the half-light radius of the UCD. 
This means that our measured velocity dispersion is intermediate between the central ($\sigma_{\rm 0}$) 
and global ($\sigma_{\infty}$) velocity dispersions. We therefore use the slit size (both width and length
integrated over), seeing, and half light radius and the best fitting Sersic index from the HST imaging 
(described in \citealt{Norris&Kannappan11}) to calculate the correction between aperture velocity dispersion
and infinite velocity dispersion following the approach of \citet{Strader11a}. As expected the correction
factor is small, 0.92, leading to an 18$\%$ decrease in the derived total mass over using the aperture
velocity dispersion. The final virial mass arrived at, when applying the correction is 
M$_{\rm vir}$ =2.17$^{+0.51}_{-0.51}$$\times$10$^7$ M$_\odot$. This mass is considerably larger than
the value of M$_{\rm vir}$ =1.54$^{+0.43}_{-0.43}$$\times$10$^7$ M$_\odot$ found in \citet{AIMSSII}, 
with the difference due principally to the use of an object specific choice of C for this UCD.

\subsubsection{Jeans Mass Modelling}

Given the observed flat velocity dispersion profile we do not expect that this UCD hosts a particularly
massive black hole like the one found in M60-UCD1 \citep{Seth14}. However, we have carried out
a simple Jeans analysis both to test this suspicion and to examine the remaining tension between 
the dynamical and stellar masses of this UCD.

We start by using GALFIT \citep{Peng02_galfit} to fit a Sersic profile to the available HST WFPC2 planetary 
camera F606W band imaging. This imaging had the large scale galaxy background from 
NGC~4546 removed using the approach described in \cite{Norris&Kannappan11}. The GALFIT modelling 
of the light distribution made use of a simulated TINYTIM \citep{tinytim} PSF constructed at the location of 
the UCD, as no suitable stars were located within the PC chip field of view. The resulting fit was excellent, 
yielding a half-light radius consistent with the one derived from a curve of growth analysis, a Sersic n of 
1.4, and an ellipticity very close to 1 (0.92).

These values, along with the total integrated V band magnitude of the UCD were then used to produce
a multi-gaussian expansion (MGE) model of the UCD light distribution, as a required input for a Jeans 
Anisotropic MGE \citep[JAM:][]{JAM} dynamical model of the UCD. We use the IDL implementation
of the JAM code\footnote{available at http://www-astro.physics.ox.ac.uk/~mxc/software/} and fit a series 
of simple spherical Jeans models to our measured resolved kinematics and the MGE model of the light
distribution. We fit models with M$_{\rm dyn}$/L$_{\rm V}$ varying from 0.5 to 3.5, and black hole mass from 
10$^{4}$ to 10$^{8}$ M$_\odot$.

\begin{figure} %  figure placement: here, top, bottom, or page
   \centering
   \begin{turn}{0}
   \includegraphics[scale=0.7]{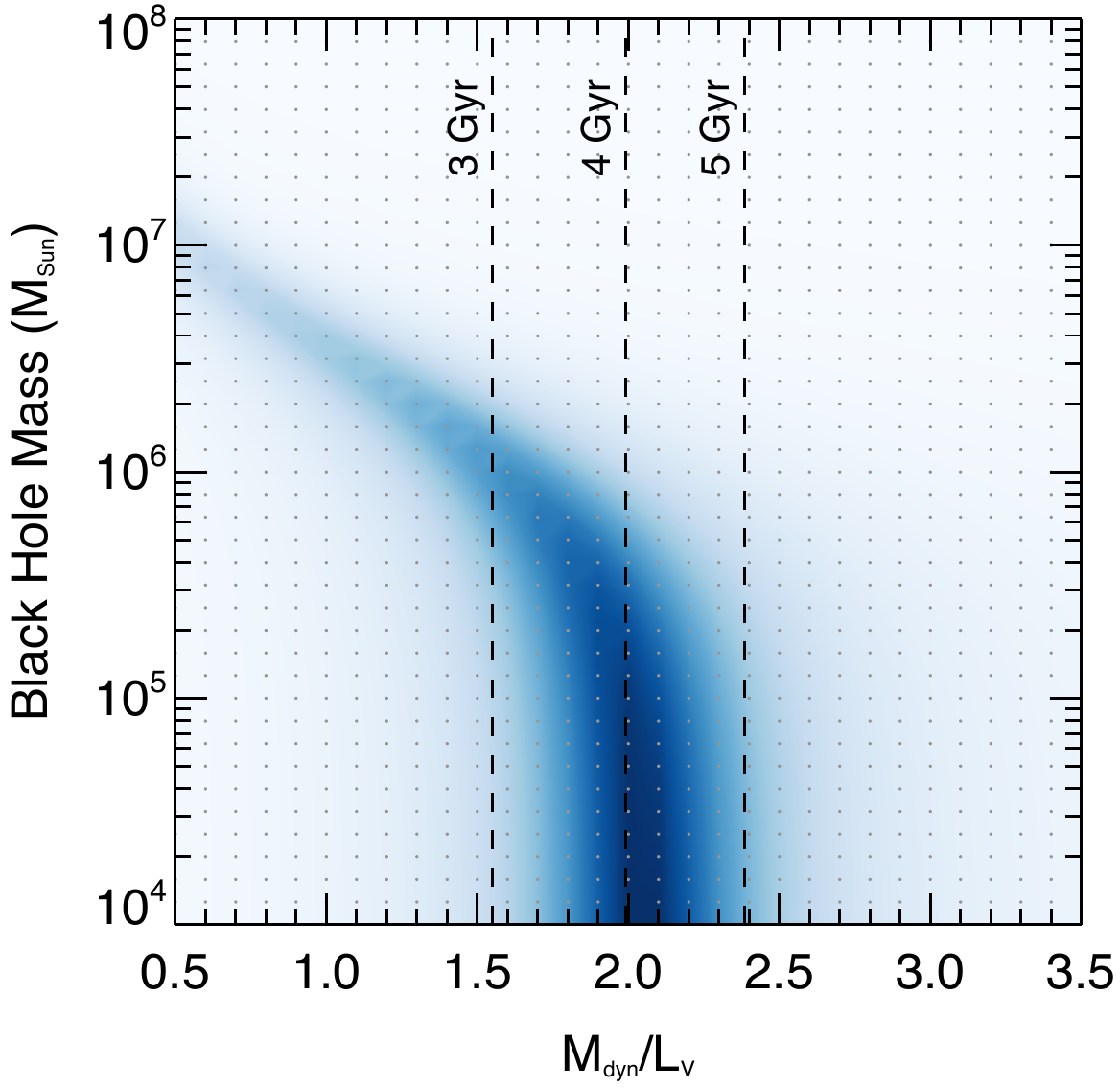}
   \end{turn} 
   \caption{The goodness of fit from the JAM modelling of NGC~4546-UCD1. The shading indicates
   the best fitting models, the small dots indicate the values of M$_{\rm dyn}$/L$_{\rm v}$ and black hole mass that were
   fit to the data. The dashed lines show the predicted mass-to-light ratios for SSP models 
   with a Kroupa IMF, [Z/H] = +0.35, and 3, 4, and 5 Gyr \citep{Maraston98,Maraston05}. }
   \label{fig:blackhole_ml}
\end{figure}

Figure \ref{fig:blackhole_ml} shows the result of this fitting procedure. As expected we are only able to 
place an upper limit on the black hole mass, the black hole can have any mass less than around 10$^{6}$ M$_\odot$,
without being in significant tension with the stellar mass-to-light ratio expected for the UCDs age and metallicity.
Furthermore, the best fit dynamical mass-to-light ratio (of 2) is indistinguishable from a purely stellar 
mass-to-light ratio for a stellar population with the same parameters as NGC~4546-UCD1, indicating that 
there is little sign of a significant dark mass component within this UCD, be it a black hole or dark matter.
The result of the fitting does however provide additional confirmation that the dynamical mass of the UCD 
is consistent with the stellar mass of the UCD only when the UCD is relatively young.

It is worth noting that assuming that the UCD mass is the same as the original central NSC, 
a black hole of sufficient mass to be detectable is not necessarily expected. Using the relation between central 
massive object (either NSC or super massive black hole) and total dynamical galaxy mass from \citet{Ferrarese06} 
implies that the original host galaxy of the UCD should have had a dynamical mass of 1-2 $\times$10$^{10}$ M$_\odot$
(and therefore a total stellar mass of several $\times$10$^{9}$ M$_\odot$). 
This mass is similar to that of M33 \citep{Corbelli03}, the local group spiral galaxy which also hosts a NSC
with very similar velocity dispersion to that of NGC~4546-UCD1 ($\sigma_{e}$ = 24 $\pm$ 1.2 km/s), but has no 
detectable black hole down to a upper limit of 1500 M$_\odot$ \citep{Gebhardt01}. Similarly, the NSC of
NGC~404 which has a similar dynamical mass as NGC~4546-UCD1 (1.1$\times$10$^7$ M$_\odot$), also 
lacks a black hole more massive than 5$\times$10$^5$ M$_\odot$ \citep{Seth10}. In fact, apart from 
M60-UCD1 no object with a mass $<$ 10$^{\rm 10}$ M$_\odot$ is known to host a black hole more massive
than 10$^{\rm 6}$ M$_\odot$ \citep{Miller15}.

The fact that we find the most likely black hole masses to be consistent with those of SMBHs found in 
galaxies similar to the expected progenitor galaxy implies that while the progenitor galaxy was destroyed 
the central bound structure that was left to evolve into NGC~4546-UCD1 cannot itself have been heavily 
stripped. If the opposite was true, and the UCD itself was significantly stripped, we would likely find a 
black hole significantly over-massive relative to the expected mass, as is the case for M60-UCD1. 
This apparently low black hole mass also supports the idea that the progenitor of NGC~4546-UCD1 
was a small dwarf galaxy with a nuclear star cluster, not a genuine bulge. This is in contrast to the situation for 
M60-UCD1, where the black hole is clearly detected and has a mass of 2.1$\times$10$^{7}$ M$_\odot$ within 
a UCD of only 1.2$\times$10$^{8}$ M$_\odot$. Using the same scaling relations implies that original central 
bound structure mass of M60-UCD1 was around 100 times larger than is left in the UCD, and at a mass of 
$\sim$ 10$^{\rm 10}$ M$_\odot$ this structure was clearly a genuine bulge. Further evidence for this picture 
is provided by the observed two component structure of M60-UCD1, with one of the structures having properties 
consistent with being a nuclear star cluster embedded within the remnants of the mostly stripped bulge \citep{Seth14}.

Under the assumption that NGC4546-UCD1 either never contained any dark matter, or has been so thoroughly 
stripped that it has lost any significant dark matter component within the region our measured dynamics probe, we 
can also make an estimate of the total dynamical mass using the best fit M$_{\rm dyn}$/L$_{\rm V}$. This is because 
under these assumptions M$_{\rm dyn}$/L$_{\rm V}$ = M$_\star$/L$_{\rm V}$. The best fit M$_{\rm dyn}$/L$_{\rm V}$ 
was 2.0, which when using the total V band magnitude of the UCD listed in Table \ref{tab:properties}, translates into 
a total dynamical mass of M$_{\rm dyn}$ = 2.59$^{+0.49}_{-0.49}$$\times$10$^7$ M$_\odot$.

The two estimates of dynamical mass are in good agreement, and furthermore are in reasonable agreement with 
the stellar masses derived in Section \ref{Sec:stellarmass}. In fact, when considering their mutual errors the dynamical 
mass to stellar mass ratio of NGC~4546-UCD1 is now consistent with being 1. A similar finding of consistent dynamical 
and stellar mass estimates was also found for a sample of nine NSCs by \citet{Walcher06}, where the ratio of the M/L 
derived from dynamics and from stellar populations were found to be consistent with 1. This is further proof that the 
dynamics of this UCD (and even NSCs) can be explained without the need to invoke exotic initial mass functions or 
the presence of dark matter or massive dark objects like black holes.

\section{NGC~4546-UCD1's Galactic Origin}

We have examined the spatially resolved kinematics and star formation history
of NGC~4546-UCD1, a UCD suspected of being the result of a tidal interaction between NGC~4546
and a dwarf companion of mass around 3 $\times$10$^{9}$ M$_\odot$ (\citealt{Norris&Kannappan11}
based on the observed relation between the nuclear mass of a galaxy and its spheroid from \citealt{Graham09}).

Our observation that NGC~4546-UCD1 has disordered kinematics is readily understandable in such a 
scenario, as nuclear star clusters are often observed to have such complex kinematic structures
\citep{Seth10,Lyubenova13}. In this context these structures are thought to result from either the 
accretion of fresh gas and subsequent star formation in-situ in the NSC \citep[as seen for example in 
the Milky Way NSC:][]{Bartko10,Pfuhl11}, or alternatively from the accretion of entire star clusters formed 
externally which have in-spiralled into the NSC from larger radii due to tidal friction \citep[see e.g.][]{Antonini12,Antonini13}.

A more conclusive indicator of the galactic origin of NGC~4546-UCD1 is provided by the temporally extended star
formation history of this UCD. The very extended nature of the star formation, which appears to 
have extended from the earliest epochs until a few Gyr ago, conclusively rules out a star cluster 
origin for this object. This interpretation is further confirmed by the star formation history of the suspected 
GC-type UCD NGC~3923-UCD1, which is entirely consistent with forming in a single-metallicity burst 
$\sim$ 10 Gyr ago, indicating the soundness of our derived star formation histories.

The fact that there does not appear to be significant metallicity evolution during the period where NGC~4546-UCD1 
was forming stars indicates that the proto-UCD was constantly accreting fresh gas (as expected if the star 
formation was to continue for such an extended period). This observation fits naturally with a picture where 
a nuclear star cluster accretes fresh lower metallicity gas from the galaxy's disc, or alternatively accretes star 
clusters which have inspiralled from the disc. The high absolute value of the metallicity also 
supports an origin of this object as the central regions of a relatively massive galaxy, as only massive galaxies
are able to enrich to such extreme metallicities. The fact that our inferred galaxy mass is lower than that expected
for a galaxy with such high metallicity perhaps indicates an emerging tension with the very high metallicities observed
in known stripped nucleus UCDs (i.e. M60-UCD1 also has [Z/H]$\sim$ +0.2). Examination of a larger sample of 
confirmed stripped nuclei will be required to address this tension further. 

We can now sketch out a complete picture of the history of NGC~4546-UCD1 that starts with it forming
in the core of a dwarf galaxy at early epochs. For the next 10+ Gyr it continued to form stars at a low rate,
with occasional acquisition of lower metallicity gas or already formed stars (in the form of inspiralling star
clusters) from the galaxy's disc. Then some point in the last 1-3 Gyr it began a final interaction with
NGC~4546 which led to the destruction of the main body of the dwarf galaxy, and hence cut the
nucleus off from a supply of fresh gas or stars. The exact timing of the interaction is difficult to judge as 
the errors on the youngest age component are not inconsiderable, and there is also the possibility that
star formation continued for some period after the stripping event, using up gas already present in the
nuclear regions (its also possible the interaction funnelled more gas to the nucleus). One way to more
precisely date the interaction would be if some bona-fide GCs were formed from material from the dwarf 
during the merger event, and now survive mixed in amongst the previously existing GCs of NGC~4546.

\section{Conclusions}

We have used deep spectroscopic observations of NGC~4546-UCD1, one of the closest ultra compact 
dwarfs to examine its stellar populations and star formation history. We find evidence that the internal 
kinematics of this object are complex, and possibly indicative of multiple periods of star formation. This 
suggestion is confirmed by our analysis of the star formation history of this object, which confirms that
NGC~4546-UCD1 was forming stars from the earliest epochs until 1-2 Gyr ago. Such a star formation 
history is not expected in a star cluster and can only be found in true galactic populations, thereby 
confirming that this UCD is the remnant nucleus of a galaxy tidally disrupted by NGC~4546. Comparison 
with the star formation history of another UCD, NGC~3923-UCD1, that is suspected of being a massive 
star cluster, demonstrates that the extended star formation history of NGC~4546-UCD1 is a robust detection, 
and further supports the contention that UCDs are a composite population of massive star clusters and 
stripped galaxy nuclei.

We suggest that similar spectroscopic observations of other UCDs may provide a more observationally 
straightforward method (compared to AO assisted kinematic studies) to detect objects formed by tidal 
interactions.

\section{Acknowledgements}

The authors would like to thank the referee, whose very constructive comments
significantly improved this work. In addition, the authors would like to thank Alexandre Vazdekis 
for providing the latest version of his stellar population synthesis models prior to public release. 
We would also like to thank Glenn van de Ven and Eva Schinnerer for their very helpful discussions 
which greatly improved this manuscript, and Jay Strader for computing the correction 
factors used to correct measured aperture velocity dispersion to total velocity dispersion.

CGE and FRF acknowledge financial support from Consejo Nacional de Investigaciones Cient\'ificas 
y T\'ecnicas (PIP 0393), and Universidad Nacional de La Plata, Argentina (G128).

Based on observations obtained at the Gemini Observatory, as part of programs GS-2011A-Q-13, 
GS-2013A-Q-26, GS-2014A-Q-30, and processed using the Gemini IRAF package. The Gemini 
Observatory is operated by the Association of Universities for Research in Astronomy, 
Inc., under a cooperative agreement with the NSF on behalf of the Gemini partnership: the 
National Science Foundation (United States), the National Research Council (Canada), 
CONICYT (Chile), the Australian Research Council (Australia), Minist\'{e}rio da Ci\^{e}ncia, 
Tecnologia e Inova\c{c}\~{a}o (Brazil) and Ministerio de Ciencia, Tecnolog\'{i}a e Innovaci\'{o}n 
Productiva (Argentina).

Support for Program number HST-AR-12147.01-A was provided by NASA through a grant 
from the Space Telescope Science Institute, which is operated by the Association of Universities 
for Research in Astronomy, Incorporated, under NASA contract NAS5-26555.

This work is based in part on observations made with the Spitzer Space Telescope, which is 
operated by the Jet Propulsion Laboratory, California Institute of Technology under a contract with NASA.

This research has made use of the NASA/IPAC Extragalactic Database (NED) which is operated 
by the Jet Propulsion Laboratory, California Institute of Technology, under contract with the 
National Aeronautics and Space Administration.

\bibliographystyle{mnras}
\bibliography{references}

\appendix

\label{lastpage}

\end{document}